# An Empirical Study on Technical Debt in a Finnish SME

Valentina Lenarduzzi
*Tampere University*
Tampere, Finland
valentina.lenarduzzi@tuni.fi

Teemu Orava

Tampere, Finland
teemu.orava@kapsi.fi

Nyyti Saarimäki
*Tampere University*
Tampere, Finland
kari.systa@tuni.fi

Kari Systä
*Tampere University*
Tampere, Finland
kari.systa@tuni.fi

Davide Taibi
*Tampere University*
Tampere, Finland
davide.taibi@tuni.fi

*Abstract*—*Background.* **The need to release our products under tough time constraints has required us to take shortcuts during the implementation of our products and to postpone the correct implementation, thereby accumulating Technical Debt.**
*Objective.* **In this work, we report the experience of a Finnish SME in managing Technical Debt (TD), investigating the most common types of TD they faced in the past, their causes, and their effects.**
*Method.* **We set up a focus group in the case-company, involving different roles.**
*Results.* **The results showed that the most significant TD in the company stems from disagreements with the supplier and lack of test automation. Specification and test TD are the most significant types of TD. Budget and time constraints were identified as the most important root causes of TD.**
*Conclusion.* **TD occurs when time or budget is limited or the amount of work are not understood properly. However, not all postponed activities generated "debt". Sometimes the accumulation of TD helped meet deadlines without a major impact, while in other cases the cost for repaying the TD was much higher than the benefits. From this study, we learned that learning, careful estimations, and continuous improvement could be good strategies to mitigate TD These strategies include iterative validation with customers, efficient communication with stakeholders, meta-cognition in estimations, and value orientation in budgeting and scheduling.**

## I. INTRODUCTION

Companies commonly spend time to reduce the Technical Debt (TD) in their systems. Many factors can lead to TD; they can be internal, related to the business or the environment, or they can be external to the company [1].

TD is a metaphor from the economic domain that "refers to different software maintenance activities that are postponed in favor of the development of new features in order to get short-term payoff" [2].

Technical issues include any kind of information that can be derived from the source code and from the software process, such as usage of specific patterns, compliance with coding or documentation conventions, architectural issues, and many others. For example, when a new feature does not fit the current architecture, the incompatibility might solved with an immature implementation [2] than will be fixed in the future implementing a proper solution.

Researchers have investigated different aspects of TD and proposed different approaches to pay it back. However, only few works have investigated concrete cases and identified the root causes of TD in companies.

In this work, we report on an empirical study we performed in our case company, a Finnish SME that operates in Business-to-Business sector and develops web applications for managing sales channels.

We identified cases where we postponed different activities and then analyzed the reason(s) for the postponement, the issues generated by the postponement, and how the post- poned activities were implemented later. We also highlight the overhead generated by the postponement of the activities themselves (the interest).

The results of this work can be beneficial not only for the scientific community but also for other companies. As other companies can understand the reasons why our case company postponed some activities, and the issues generated by the postponement, they can make more informed decisions in similar situations. The results of this work confirm that TD can cause significant economic losses if payback is postponed. Also, postponing activities - even if it is beneficial in the short term - can often be an economic disadvantage.

We investigated our case company's TD with a focus group involving five members of the company. Our main goal was not to regret past losses, but to understand what happened in the past and find ways to prevent similar situations.

The remainder of this paper is structured as follows. Section II reports on related work. In Section III, we introduce the empirical study design and report the results in Section IV. The discussion is presented in Section V and conclusions are drawn in Section VI.

## II. RELATED WORK

There are several ways to prevent TD. For example, Fowler suggested that software should be designed to be strangled, i.e., to be surpassed by new versions easily [3], while according to Cunningham, utilizing the modularity of objects allows developing flexible software [2]. However, sometimes debt cannot be avoided and in order to avoid rising costs, the generated debt should be paid back as soon as possible.

According to Z. Li, TD occurs when technical shortcuts are taken to gain short-term benefits that are harmful for the system in the long term [4]. There are several reasons that lead to technical compromises, such as unrealistic schedule, budget constraints, or estimation errors. Highly indebted products become inflexible and unprofitable, and the accumulation of debt eventually leads to dead end whereupon the system has to be replaced with a new one.

Klinger et al. [5] interviewed four software architects to understand how decision-making regarding TD was conducted in an enterprise environment. The results showed that the decisions related to TD issues were often informal and ad-hoc, which prevented tracking and quantifying the decisions and issues. Moreover, just as in our work, this study also reported that there was a large communication gap between technical and business stakeholders in the discussions related to TD.

Ampatzoglou et al. [6] conducted a multiple case study in the embedded systems industry in order to investigate the expected lifetime of components affected by TD and the most frequently occurring types of TD. They considered seven embedded systems industries from five different countries. The results showed that in order to increase the expected lifetime of components, maintainability plays a major role. Moreover, they found the most frequent types of TD to be test, architecture, and code.

Recently, De Toledo et al. [7] conducted an exploratory case study with a large company on a project with about 1,000 services. They investigated Architecture TD in the communication layer. The study combined an analysis of existing documentation and interviews to identify issues, solutions, and risks, providing a list of architectural issues that generate TD.

## III. FOCUS GROUP

In this section, we describe the design of our study, including the goal, the research questions, the study context and procedure, and the data analysis.

The case product was a sales channel management tool that the case company offers as a service (Saas). The company is a micro-enterprise (less than 10 person), that develop a single product (the sales management tool). The product is customized for suppliers, providing a limited set of features, depending on their needs.

The product has been developed for 4 years (from January 2015) and it is based on JavaScript and NoSQL and it's developed with the MEAN stack (MongoDB, Express.js, AngularJS and Node.js).

### A. Research Questions

Based on the aforementioned goal, we derived the following Research Questions (RQs):

- **RQ1**: What are the most common types of TD?
- **RQ2**: What are the main causes of the accumulated TD?
- **RQ3**: How to mitigate TD?

**RQ1** aims to determine the most common types of TD in the company and their impact on business.

**RQ2** aims to investigate the causes of the TD identified in the company.

**RQ3** aims to identify ways to prevent TD from occurring in the future based on the knowledge gained by RQ1.

### B. Planning the study

We planned a focus group to last from two to three hours. We identified a number of issues to be covered that were sufficient for having a meaningful discussion and interaction between the participants.

We selected five participants: the Chief Technology Officer (CTO), the Chief Financial Officer (CFO), the Chief Marketing Officer (CMO), and two developers. All participants voluntarily participated in the study, as they were interested in how to avoid facing similar situations as in the past and wanted to understand which activities should not be postponed.

The session was moderated by one of the authors, that did not vote nor participated on the identification of TD. Before the session, the moderator introduced the goals and the rules of the focus group. Then he presented the following six discussion topics:

T1: Which activities have been postponed in the past?
  This topic was investigated in two steps: First, the participants answered this question individually, reporting the activities on post-it notes. Then the moderator asked them to read their list of activities and grouped the same activities on the whiteboard.

T2: Which type of TD was generated by the postponed activities?
  The participants grouped the postponed activities based on the type of TD. We adopted a classification of eleven categories, including the ten TD categories proposed by Li et al. [4] (Requirement TD, Architectural TD, Design TD, Code TD, Test TD, Build TD, Documentation TD, Infrastructure TD, Versioning TD, Defect TD) and one new category (Organizational TD).

T3: What were the reasons for postponement?
  Regarding this topic, the participants discussed the motivations for the postponement of the activities and then reported them on the activity post-it notes created in T1.

T4: How were the activities then implemented? In this task, the participants reported the solutions adopted to implement the postponed activities and reported them on the activity post-it notes.

T5: What problems did the postponement cause?
  The participants collaboratively discussed the problems that caused the postponement, including economic, technical, and organizational ones. In this case as well, they reported them on the activity post-it notes.

T6: Ranking the importance of the problems.
  In this task, each participant received ten votes, in the form of adhesive "dots", and was asked to vote on the most harmful problems. The participants were free to distribute their votes as they liked, for example, assigning all ten votes only to one activity or distributing them evenly among the activities.

Except for Topic 1, if needed, participants were allowed to use extra post-it notes connected to the same activity.

*C. Data Analysis*

The data collected from the focus group was analyzed by determining the proportion of responses in each category. TD causes were analyzed following the 5-Whys technique [8].

## IV. RESULT

In this section, we will first report the perceived TD issues highlighted by the participants, together with the problems the issues generated and their causes. Finally, we will answer our research questions.

*A. Perceived Debt*

The participants if of focus group identified 10 different types of TD (TD items).

*Organizational and Product Management Issues*

TD1 *Implementation of multiple versions of the same product, as different customers wanted to use the system for different purposes. (Requirements TD, Organizational TD)* The prioritization of the features and tasks as well as the estimation of the cost and other effects of the customer-specific tailoring became difficult. The recognized causes where Specification issues, Budget constraints, Estimation issues and Time constraints (e.g. related to Fast Delivery).

TD2 *Disagreement with supplier about the Minimum Viable Product (MVP) [9]. (Requirements TD)* The first version of the system was subcontracted from an external vendor that wanted to implement the initially agreed specification instead of iterative development and adapting to improved understanding of the customer needs. The recognized causes where Specification issues, Budget constraints and Estimation issues.

*Architectural Issues*

TD3 *Lack of multitenancy causes budgeting increase and lack of flexibility (Infrastructure TD).* The products are delivered as SaaS services, but the implementation forces a totally separated installation for each customer. This raises the operation and infrastructure costs. Multitenancy was not originally the highest priority and then the need of introducing it is costly. The recognized cause was Budget constraints.

TD4 *Hard to maintain a simple User eXperience (UX) with the growth of functionalities. (Design TD)* The UX was designed by the supplier that did not want to redesign it anymore, creating issues in adding new features while maintaining a good user experience. The recognized cause was Time constraints.

*Development and Testing issues*

TD5 *Lack of automatic testing costs more in the future* (Infrastructure TD). The testing budget was too low to enable the creation of automatic testing during development since the company did not even have enough time to concentrate on fast delivery to the client. The recognized cause were Time constraints.

TD6 *Testing is expensive. (Test TD)* The company lacked dedicated tested and had human resourcing challenges. The focus group was not able to find the actual cause of this TD.

TD7 *Low code coverage in tests causes risks in development and additional work. (Test TD)* It was hard to estimate budget and schedule in the beginning and the company had to postpone some testing. Also, the company did not have dedicated personnel for testing, and developers were not as efficient in testing as dedicated tester would be. The recognized causes were Estimation issues, Communication issue and Budget constraints.

*Source Code Maturity Issues*

TD8 *Lack of code documentation. (Documentation TD)* The case company was commonly too busy to create code documentation as new features has usually highest priorities. The recognized cause was Time constraints.

TD9 *Technical shortcuts (Code TD)* These TD items are present due to lacking time and budget. The recognized causes were Time constraints and Budget constraints.

TD10 *Duplicated code (Code TD)* Developers failed to follow the "Don't Repeat Yourself" principle and modularize the implementations. Instead they duplicated the code because they were in hurry. In some case, the company had no time to extend or generalize the existing code. The focus group was not able to find the actual cause of this TD.

*RQ1. What are the most common types of TD?*

The focus group considered the Test and Requirements TD as clearly more significant than other types of TD, as reported in Table 1a.

*RQ2. What are the main causes of the accumulated TD?*

The causes of the perceived TD items are summarized in Table 1c

1) Budget constraints (TD1, TD2, TD3, TD7, TD9) and time constraints (TD1, TD4, TD5, TD8, TD9) are the most recurring reasons. Estimation issues (TD1, TD2, TD7) is also a significant cause and closely related to budgeting and timing.
2) Time-related causes (Time constraints), usually related to fast delivery, recurred almost as frequently as budget constraints. It can be speculated that the lack of time depends on the budget.
3) Other causes were not as significant.
4) In some cases, the causes of the TD remain unknown.

*RQ3. How to mitigate TD?*

Based on the discussion of the focus group we highlighted three main aspects that could be improved to mitigate TD.

1) *Learning from customers.* Organizations have to understand what should be built using prototypes and validation with customers.

TABLE I: Results. Motivations are counted once for each TD

(a) Perceived by interviewees and total points

| TD Description | Points |
|---|---|
| TD2. Disagreement with supplier | 7 |
| TD5. Lack of automatic testing | 7 |
| TD1. Implementing multiple products | 3 |
| TD0. Technical shortcuts | 3 |
| TD6. Expensive tests | 2 |
| TD3. Lack of multitenancy | 2 |
| TD8. Lack of code documentation | 2 |
| TD7. Low code coverage in tests | 2 |
| TD10. Duplicate code | 2 |
| TD4. Hard to maintain simple UX | 0 |

(b) perceived TD types and sum of point

| TD Type | Points |
|---|---|
| Test TD | 11 |
| Requirements TD | 10 |
| Code TD | 5 |
| Organizational TD | 3 |
| Infrastructure TD | 2 |
| Documentation TD | 2 |
| Design TD | 0 |
| Architectural TD | 0 |
| Build TD | 0 |
| Versioning TD | 0 |
| Defect TD | 0 |
| Grand Total | 30 |

(c) Count of TD motivations presented

| Possible cause of motivations | Count |
|---|---|
| Budget constraints | 5 |
| Time constraints | 5 |
| Estimation issues | 3 |
| Specification issues | 2 |
| Communication issues | 1 |
| Design issues | 1 |

2) *Careful estimation.* The whole organization should understand the technical boundaries to avoid estimation errors. They should use previous tasks to improve their effort estimation regarding the development of new tasks. Underestimation can cause additional expenses for company. Customers should pay for the overall costs of the system; they tend to pay only for visible costs, which are only the tip of the iceberg. The costs of testing and documentation, which tend to be under the surface, should be made visible to them. The company has to find the right pricing balance in order to remain competitive. Underestimating the amount of work can lead to compromises in less visible costs.

3) *Continuous improvement.* Organizations can gradually improve the quality. Deficiencies in development areas should not be postponed. Companies should invest in testing and documentation because their TDs are hindering development and ultimately take up a lot of the developers' precious time. A lack of tests increases the need for manual testing and the risk of regression. Lack of documentation diminishes knowledge and adds tacit knowledge. Evanescence of knowledge will accumulate the costs of testing and documenting over time. Companies have to find the critical point in mitigating TD where benefit is bigger than cost.

## V. DISCUSSION

Identifying TD and its possible root causes helped the company to understand their most critical issues. Conversation helped to determine the causes of accrued TD to enable mitigating TD in the future. Ways to mitigate TD were explored based on the results. Budget constraints were considered as the most critical root cause of TD; however, time constraints and fast delivery were considered almost as critical.

Time constraints can be related to budget constraints when they are caused by HR constraints. However, they do not always relate to budget as more employees do not automatically remove time constraints. According to FP. Brooks [10], work distribution follows Amdahl's law. Thus, the more work is distributable, the more time is saved by adding developers. The required learning curve and the need for more communication lessen the benefit of having more employees. Thus, even if the budget is sufficient, time constraints can remain until the team reaches its optimal performance in group development, as reported by B. Tuckmann [11].

When discussing the lack of documentation, the CTO said, *"We faced this TD about not having the documentation when you [developer] came and we did these bug fixes during the autumn. Had there been these, I think it would have been a little bit easier."*

### A. Learning from Customer

Learning from the customers is the first answer to RQ3 regarding how to mitigate TD. As stated by one developer, when the organization knows the customer's needs, it it is easier to go in the right direction. When there are many customers, User Experience Design becomes more important as a generalized solution has to satisfy everyone's needs at the same time. *"We are kind of having it done by experimenting and communicating more with the customers to understand what they need and we are doing it in an iterative way to solve the customer's problem, but this works until we have only handful of customers."*

Idea validation with users using prototyping could follow validated learning as suggested by E. Ries [12]. According to M. Christel et al. [13], the customer has to be supported in requirements elicitation because the customer's understanding is limited. However, the CTO stated that the customer should be consulted only for major decisions and should not be bothered with every minor detail.

Problems caused by a lack of validation were emphasized when the subject company outsourced their software development. Sections that are more important for the business than strategically can be outsourced. Outsourcing can allow companies to focus on their core competence, but suppliers have their own interests and all the decisions have to be well-reasoned. The CFO stated *"that was why we were so upset with them [the Supplier] because the plan was to have something not so solid in the back end but we could have a couple of customers to actually test. Problem is that they chose not to give us that; we had to wait two years before we were able to have a customer to test the MVP and that was their big mistake."*

*B. Careful Estimation*

Careful estimation is the second answer to RQ3. SMEs need to use their budget wisely. A limited budget forces a company to generate TD which it hopes to pay back as soon as possible. Payback time can be when the company gets enough funding. The risk of a roadblock through a "TD bankruptcy" increases when new requirements emerge and need attention, leading to the rewriting of existing features, which should be avoided by estimating the costs of TD. Moreover, outsourcing part of the development to consultants, also increase the risk of requirement TD [14] related to misunderstandings [15] and increase the communication overhead [16].

For an SME, budget constraints are inevitable and the company needs ways to cope with its budget. Considering the lifecycle of companies, Nilsson et al. [17] claimed that in the pre-deployment phase, architectural and structural TD should be avoided. Other types of TD such as test and documentation TD can be incurred in that phase. Communication is important especially at the beginning in order to avoid wrong decisions that later become TD. Any historical analysis of budget constraints is speculative and thus no single reason can be named. Inadequacy in in specification validation can drain the budget. TD2, disagreement with the supplier, was one of the most important TD items. The CFO stated that concrete prototypes could have helped validation. Lack of prototyping consumed time and caused bitterness. Decisions made were not optimal in the long term and caused TD.

Planning requires good communication. Stakeholders should make sure they consider every aspect of new features, utilizing, for example, the Architecture Trade-off Analysis Method (ATAM) to find possible trade-offs in architecture decisions by formally listening to all stakeholders [18]. A customer approaches the product from a top-down perspective. They cannot see all the technical details related to the implementation of a feature. On the other hand, developers are able to see the bottom-up perspective and know all the technical aspects quite well. However, they might have deficiencies in domain knowledge and cannot value all the customer's needs. Both parties become victims of the Dunning-Kruger effect [19] when they fail to look below the surface.

When discussing the implementation of features for multiple products, the CFO said *"I don't think they [the Supplier] took really that much time to understand because in every meeting we repeated the same. It was very important and in the specification, the written specification, and even in the contract they signed, this was written."* The supplier's developer for his part commented *"the Client's team was not able to convince us of that and explain the idea really well. The reason is the domain knowledge, the deficiency on the Supplier's side."*

Just as with suppliers, companies face challenges in justifying the work to be done with customers. The customers do not see the less visible costs, which should be communicated to them as they improve the long-term health of the system. The CFO gave the following example: *"We had an issue that one of our customers wanted to modify the questions. [...] It was quite a big change and they said that 'no, we won't pay that much' and then we said we cannot do it. They were not very happy but we had no choice. It was too expensive and the client did not see any value in that."*

Careful estimation avoids risks. In addition to communication issues, estimation errors can be reasons that drain the budget. As mentioned in the Results, underestimation is unprofitable for the company. Estimation errors occur because of unpredictable complexity of a task. Developers might not see all the sub-tasks concerning a new task when doing the estimation at the beginning of the development of a new task. Every new task is unique and has little in common with the previous tasks. A little knowledge of the subject makes developers overconfident, which leads to them underestimating the amount of work. Again, only the tip of the iceberg is seen and a new task is seen as simpler than it actually is in the end.

When pricing and schedule are unrealistic, development will focus only on the most critical areas. Pessimism in estimation could help to improve quality, but the challenge is to maintain competitiveness with bigger companies that have economies of scale and can use their capital to fund all the various aspects. Companies can find estimation challenging, as stated by the CFO: *"The client paid us like 10,000 euros for the customization and between our hours and what we paid to do that modification it costed us 15,000-17,000 euros. We accepted the specification but we totally did not understand how much it would actually cost and how much time it would take because it was done in a rush."*

Companies can improve in finding critical point in estimates by time meanwhile preparing for estimation errors. As stated by C. Parkinson [20], his namesake law leads to overestimation since after some point, finishing a task requires the same time regardless of the allocated time. Meanwhile, according to D. Hofstadter [21], his namesake law leads to underestimation since a task requires more time than estimated although the estimator would be aware of estimation challenges.

*C. Continuous Improvement*

Continuous improvement is the third answer to RQ3. According to a developer, the quality of the code has suffered from a lack of unit tests. Testing is needed to validate conformance. One developer stated: *"Requirements were not written anywhere and if you touch and you happen to break something it's even hard to regulate what's broken until it gets into the customer's hands."*

Companies need to find the golden mean in quality improvement, where the cost-benefit ratio is the lowest [22]. Moreover, companies should consider continuous quality monitoring approaches, instead of one-shot refactoring [23] [24].

Testing, documenting, and bug fixing are ways to reduce waste in software development. Testing and fixing bugs become more difficult over time when software entropy increases. According to the CFO, *"at the latest stage when we are going to do the automated testing, which is very important anyway, it's going to cost us quite a lot because we need to dig into the old code of the application so we need to go back."*

## VI. Conclusion

In this work, we analyzed the main reasons for Technical Debt (TD) in an SME company, the problems it created, and how the company mitigated it. We investigated what happened in the past, so as to avoid making the same mistakes again, or to make reasoned choices. Our participants considered the most significant TD items to be disagreement with suppliers and lack of test automation. The most significant TD types were Test and Requirements. Possible root causes were budget constraints, estimation and specification issues, and fast delivery. Overall, the most important root causes were considered to be budget constraints, time constraints, and estimation issues.

Attempting to build a connection to management theory helps to understand the issue of TD in depth. Based on the analysis of the results and related work, the following methods can be used to mitigate TD:

- *Learning from customers* - prototyping with the customers to find the right direction and communicating efficiently with the stakeholders;
- *Careful estimation* - using meta-cognition to learn estimation;
- *Continuous improvement* - using limited budget and time wisely to bring value.

Another result is that requirements were not validated properly at the beginning when a product was outsourced to an external supplier. Moreover, the developers underestimated the time for testing and bug fixing. As estimation errors are harmful to budgeting and scheduling, awareness of one's own competence and transparency in communication can avoid risks in the future.

This work provides an overview of the main issues related to TD in our case company. However, we are aware of different threats that may have influenced the results. Some participants might not have reported some TD issues for different reasons. The presence of the company's technical management (CTO, CFO, and CMO) could have influenced the answers of the developers. The focus group was conducted over a period of two hours, and therefore we probably have not reported all the issues that occurred during the history of the company, but only the most recent or the most significant ones.

Further studies are needed to create a stronger bond between the effects of validation and estimation on the one hand and budgeting and scheduling on the other hand. Benchmarks of our estimations with existing dataset [25] and adopting TD management tools widely used by competitors [26] [27] could be a viable solutions to mitigate this threat. Understanding these laws also requires interdisciplinary studies that combine computing, quality management, and psychology. A continuous quality management approach [24] [23], could help to prevent TD. Moreover, management studies help to develop better processes, while psychology and organizational studies can explain why estimation errors occur. Understanding root causes by looking at them through these fields will result in better knowledge of TD and help SMEs avoid pitfalls, thereby enabling them to be even more successful.